# Bidirectional phase sensitivity in holographic phototransient microscopy

Emmanuel Kotu Robertson[1], Dennis van de Lockand[1], Matz Liebel[1,a]

[1] Department of Physics and Astronomy, Vrije Universiteit Amsterdam, De Boelelaan 1100, Amsterdam 1081 HZ, The Netherlands

[a] email: m.liebel@vu.nl

## ABSTRACT

Mid-infrared photothermal microscopy combines the chemical specificity of infrared absorption with the spatial resolution of visible-light detection, but practical implementations face a persistent trade-off between forward-scattering (FWS) and backward-scattering (BWS) detection geometries. FWS provides quantitative, shape-independent phase contrast but requires two-sided optical access that is difficult to achieve in aqueous or thick samples. BWS offers convenient single-sided access, but its signals are strongly distorted by depth-dependent interference for micron-scale objects. Here we present a bidirectional femtosecond mid-infrared pump-probe holographic microscope capable of switching between FWS and BWS geometries within a single instrument, and use it to introduce and validate a new imaging modality, internal forward scattering (IFS). IFS exploits the back-reflection generated at the top surface of the mid-infrared-transparent sample substrate as an internally generated forward-scattering illumination wave, isolated from the directly backscattered field via temporal coherence gating. Using polystyrene beads on $CaF_2$ substrates in air, water, and a refractive-index-matched glycerol-water mixture, we show that IFS reproduces the signal magnitudes and temporal dynamics of true FWS measurements while retaining the mechanical simplicity and single-sided accessibility of BWS. These results establish IFS as a practical, quantitative alternative to conventional FWS and BWS geometries for photothermal, and more broadly quantitative phase, imaging, with direct relevance to single-sided imaging of biological or solvent-contained specimens.

## INTRODUCTION

Mid-infrared photothermal (MIP) microscopy is an emerging and very powerful chemically specific, label-free imaging approach. MIP converts vibrational absorption, typically in the near- to mid-infrared (MIR) spectral range, into measurable refractive index or volumetric changes[1–6]. These changes are then read-out at visible wavelengths[7]. As such, MIP microscopy delivers the best of both worlds: high spatial resolution, as enabled by observations in the visible, and the chemical specificity inherent to vibrationally-selective infrared absorption[8–10]. Many different experimental implementations of MIP have been demonstrated comprising approaches relying on point-scanning as well as widefield and holographic imaging schemes[1,2,8,11,12]. A compelling advantage of MIP microscopy is its inherent compatibility with aqueous environments[13]. As a result of the visible readout, signals can be detected as long as the MIR excitation reaches the sample before total absorption. This is a crucial advantage over traditional infrared imaging, which requires extremely thin sample chambers thus complicating the experimental design[14,15] and impacting biological samples by limiting nutrient flows. MIP microscopy promises to eliminate these aspects and is thus conceptually superior, especially for studying complex biomedical processes yet the current literature remains limited to a few proof-of-concept works.

A possible explanation for this shortcoming are the stringent requirements regarding the combination of excitation and detection geometries. While MIP microscopy has the conceptual advantages outlined above, the optics required for high resolution microscopy at visible wavelengths are not compatible with MIR-requirements[16]. Additionally, the sample of interest needs to be in contact with a MIR-transparent substrate to allow MIR-excitation prior to absorption of the MIR radiation in water. Ultimately, a conceptually elegant approach becomes experimentally cumbersome, requiring complex illumination and detection arrangements, often reverting to a configuration similar to the one used in traditional infrared imaging with the only conceptual advantage being that a thin sample cell can be replaced by a thick one thus allowing sufficient nutrient exchange.

Figure 1a schematically summarises the problem. The need for MIR-transparent substrate proximity, optical MIR-access and high resolution visible readout leaves two viable configurations: a sandwich-system or dip immersion (Figure 1a, inset). Irrespective of the specific geometry, imaging can be conducted in what we term forward-scattering (FWS) or backward-scattering (BWS) with the latter containing a complex mixture of forward- and backward-scattered light as the attenuated reflection at the sample/substrate interface also interacts with the sample. Each geometry imposes distinct constraints on sensitivity, sample compatibility, and optical design[17]. FWS is often preferable if quantitative signal information for larger or heterogeneous samples is of interest but requires two-sided optical access[17–19]. Meeting this requirement is particularly challenging in solution-based measurements due to the considerations given above. BWS detection, in contrast, is naturally compatible with single-sided access, making it attractive for thick[20], opaque, or *in vivo* samples[21,22].

However, information content can be a problem as the signal amplitude of BWS is strongly impacted by the physical size of the objects. Figure 1b summarises the problem based on Mie scattering calculations for spherical objects with varying size. BWS scattering is well suited for nanoscale structures. On the microscale, however, BWS signals suffer from complex, observation angle dependent signal-modulation which can lead to reduced sensitivity and partial loss of structural information[17,20,23]. These competing behaviours mean that the optimal detection geometry depends on sample size, scattering regime, and experimental constraints. For tiny nano objects, BWS offers advantages as the signal magnitude is enhanced compared to the illumination light. However, FWS is generally preferable if quantitative signal information for larger or heterogeneous samples is of interest[17,22].

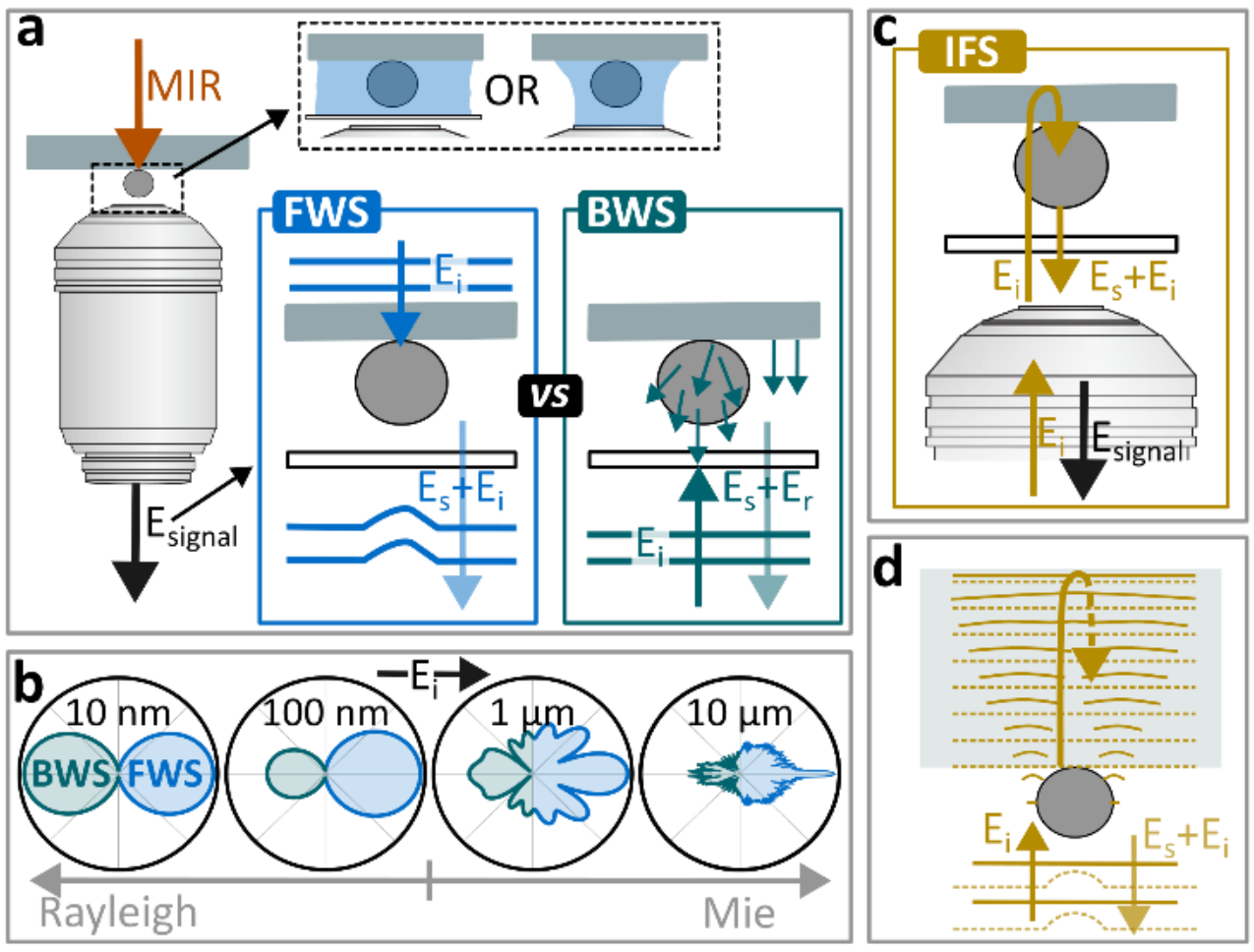

**Figure 1, Detection geometries, scattering regimes and their compatibility with mid-infrared excitation.** a) Comparison of forward-scattering (FWS) and backward-scattering (BWS) detection highlights the differences in wavefront preservation and access requirements, especially when MIR-excitation is of interest. Inset: Two

suitable sample geometries for liquid-phase specimen: sandwich and dip-immersion. b) Particle-size dependent angular scattering pattern highlighting observation angle dependent differences in information content. Simulated using MiePlot v3.0.5 with particle refractive index of 1.56 and surrounding medium refractive index of 1.0. c) Proposed internal forward scattering (IFS) geometry enabled by an internal reflection at the distal surface of the sample substrate. d) The phase information from the first pass is largely lost after the second traversal, yielding an FWS-like signal using single-sided access.

MIR-compatible substrates, objectives, and beam paths often limit the ability to switch between FWS and BWS within the same instrument. While several studies have compared FWS and BWS signals or have enabled simultaneous acquisition in both directions, these investigations have primarily focused on scattering intensity, structural contrast, or complementary imaging information rather than photothermal phase measurements. For example, recent bidirectional scattering approaches simultaneously acquire forward- and backward-scattered fields and analyse their correlations, but do not systematically evaluate geometry-dependent photothermal phase responses under identical excitation and sample conditions[17,20]. Consequently, the influence of detection geometry on photothermal phase sensitivity, to our knowledge, remains largely unexplored.

Here, we rigorously address this shortcoming, enabled by a specifically designed ultrafast implementation of MIP: a MIR phototransient holographic microscope[24–26] capable of operating in both FWS and BWS geometries. Based on the insight gained, we then propose and validate an alternative imaging modality, which we term internal forward scattering (IFS). IFS retains the experimentally advantageous single-sided optical access of BWS, while offering sample compatibility comparable to that of FWS, without requiring the complex transmission geometry typically associated with FWS. Figure 1c schematically outlines IFS which uses the *in situ* generate strong back reflection at the distal surface  of the MIR-transparent substrate, the surface opposite the specimen, to image the sample. This wave can be regarded as a quasi-plane wave even if the field acquires sample-dependent phase shifts during its first pass. The considerable distance between the surface where the sample is located and distal surface of the substrate which generates the back reflection acts as an optical low-pass filter (Figure 1d, Supplementary Information 1). Combined with coherence gated holography (Supplementary Information 2), discussed in detail below, this modality offers a practical route to obtaining images that are nominally identical to FWS-detection while preserving the mechanical simplicity and MIR-compatibility of reflection-mode operation.

The manuscript is organised as follows, we will first introduce and summarise the experimental implementation of the bidirectional phototransient microscope including important aspects related to temporal coherence gating. We will then explain the hologram and data processing pipeline for all three detection geometries. Using polystyrene (PS) beads on $CaF_2$ substrates, we then quantitatively compare the phototransient responses in FWS, BWS, and IFS for samples contained in both air and water. Finally, we conclude by measuring PS beads contained in a fully refractive-index-matched environment as an approximation of imaging objects in true 3D environments where no sample-substrate interface is present. Pump–probe temporal dynamics are provided throughout to experimentally highlight the strengths and limitations of each modality, thus providing a rigorous framework for designing optimal geometries, especially for applications where sample geometry and optical accessibility impose competing constraints.

## RESULTS AND DISCUSSION

Figure 2a schematically summarises our bidirectional MIR phototransient holographic microscope capable of operating in FWS and BWS geometries (Figure 2a). At its core, the setup can be regarded as a conventional off-axis holographic microscope, operated with 515 nm femtosecond light (Methods), which directly accesses the complex optical field of the image, including its phase[24]. In brief, a beamsplitter splits the visible input light into an illumination, or probe, and a reference field. The former interacts with the sample thus generating the signal field which contains sample-information in line with the illumination direction-dependent information transfer, as discussed in Figure 1. The signal field is collected by the microscope objective and imaged onto a camera where it interferes with the reference wave, thus generating the off-axis hologram.

Three additional features are crucial for the specific aspects of this paper, as highlighted in Figure 2: i) a flip mount which allows changing between FWS and BWS/IFS geometries without changing sample or MIR excitation; ii) a MIR-pump (Methods), coupled into the system through a Ge window, including a motorised delay line for time-resolved phototransient imaging (Figure 2b, Methods); and iii) a delay line for the reference wave which selects interference terms of interest through temporal coherence gating (Figure 2c, Supplementary Information 2). Images are acquired at 1 kHz and an exposure time of 40 µs, while modulating the MIR pump at 500 Hz to generate alternating pump-on (“hot”) and pump-off (“cold”) holograms. MIP contrast is obtained as the phase difference, or differential signal, between hot and cold frames (Methods), which reflects MIR-induced refractive-index and/or thermoelastic changes in the specimen[6,26].

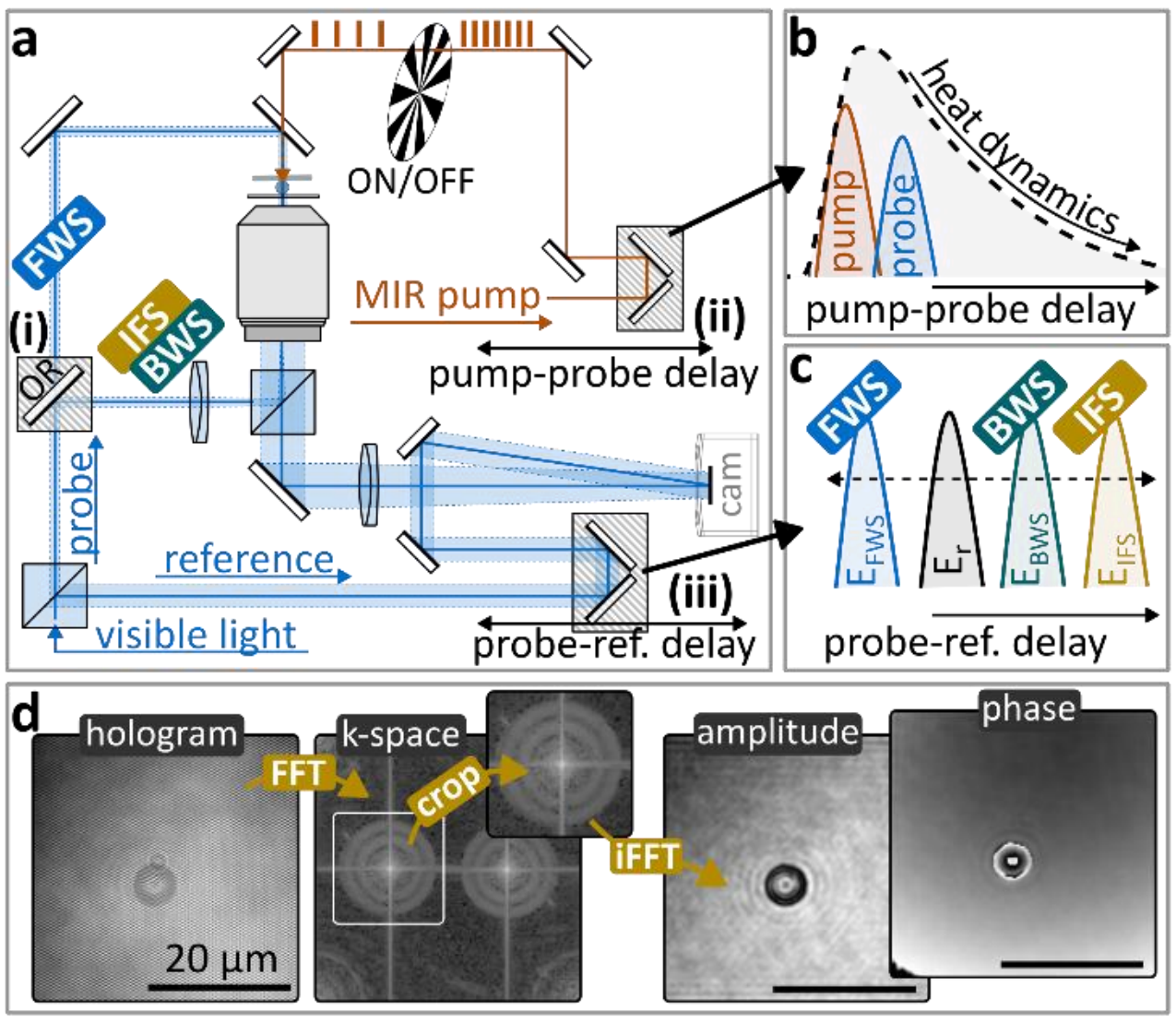

**Figure 2. Bidirectional off-axis holographic microscope design, implementation and data processing.** a) Bidirectional MIR phototransient holographic microscope design with removable pick-off (i) to switch between FWS and IFS/BWS geometries, a delay line to acquire time-resolved sample response (ii) and a delay line to adjust the probe to reference time delay (iii). b) The pump-probe delay line accesses temporal heating dynamics. c) Tuning the probe-reference delay ensures selective interference with a signal modality of interest through temporal coherence gating (see Supplementary Information 2). d) Fourier-filtering based hologram processing extracts the phase and amplitude information of interest.

Figure 2d outlines the image processing steps, from as-recorded hologram to complex field using a 5 µm diameter PS bead on $CaF_2$ in air as an example. In brief, the hologram is Fourier transformed and the momentum-shifted interference term of interest is hard-aperture selected and then isolated in image space through an inverse Fourier transformation. Phase and amplitude images are directly obtained from the resulting complex field. The division of “hot” and “cold” complex fields yields the differential phase thus accessing MIR-pump induced signals as outlined in Figure 3a.

To highlight the impact of the detection geometry on the information content of the image, it is instructive to directly compare phase and differential phase images obtained for FWS and BWS geometries. Figures 3b provides this information using the 5 µm diameter PS bead on $CaF_2$ system. In FWS, we observed a geometric appearance reminiscent of a spherical object, in line with expectations albeit a smaller than expected phase change with aliasing towards the edges, aspects we attribute to the strong phase retardation caused by the object. In contrast, the BWS images showed a sudden phase jump without clear relation to the underlying spherical geometry. These features likely caused by the complex Mie scattering patterns discussed in Figure 1b as the approximately 30 µm temporal coherence length of the probe is sufficient to capture the full back-scattered signal. Using the same field of observation, we switched from BWS to IFS detection via coherence gating, as discussed in Figure 2c. The temporal coherence length of our observation light allows conveniently isolating either the $E_{IFS}$ or $E_{BWS}$ field by adjusting the relative path length difference between signal and reference fields (Supplementary Information 2). We observed strong similarities between FWS and IFS, both for the phase and differential phase images, as also confirmed by additional through-focus scans for complementary samples (Supplementary Information 3). The minor shape differences between the two are likely due to slight illumination k-vector differences as the IFS experiment had to be performed with $k \neq 0$ to prevent parasitic back-reflections from saturating the camera.

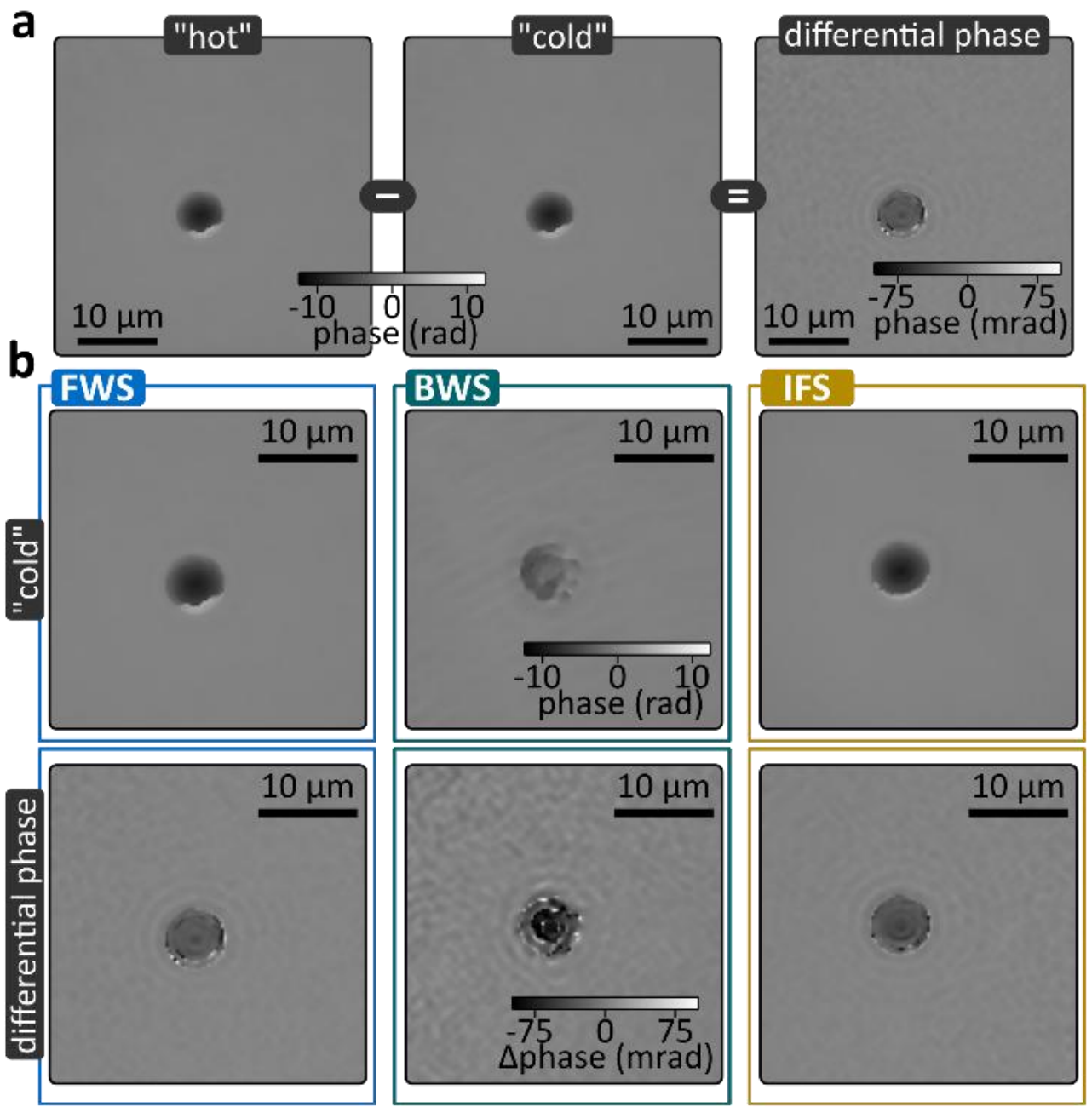

**Figure 3, Differential phase retrieval and comparison between detection geometries.** a) The differential phase is obtained as the phase difference, or complex division, between a "hot" complex image, recorded in the presence of the pump pulse, and "cold" complex image, obtained while blocking the pump pulse. b) Direct comparison of conventional, "cold", phase images and differential phase images for the three detection geometries: FWS, BWS and IFS. All images show 5 µm PS beads on $CaF_2$ in air at 8.27 ns time delay.

While static pictures provide interesting insight regarding the different modalities and a clear validation of IFS, they do not provide a full temporal picture where signal modulations through acoustic modes might play an important role. Towards this goal, we recorded differential phase images while systematically varying the pump–probe time delays. Representative time-resolved images are shown in Figure 4a for PS in air on $CaF_2$ (Methods). Irrespective of the illumination geometry, we observed oscillatory phase dynamics following MIR excitation, with comparable oscillation frequencies and gradually decreasing amplitudes, in line with our recent work on overtone-excitation based imaging[26]. Figure 4b highlights these coherent dynamics by averaging the differential phase signal over the bead area. Leaving image-quality considerations (Figure 3) aside, the consistent appearance across all three modalities confirms the integrity of the measured signals.

Figure 4a,b indicate that the temporal dynamics probed via all three imaging geometries is comparable yet signal magnitudes differ. The mean photothermal phase change is largest in BWS, consistent with the expected enhancement arising from the fact that the probe effectively passes the bead twice in a temporal window that is shorter than the coherence gating capability of our system. The illumination first passes through the bead and is then partially reflected on the first interface of the $CaF_2$ substrate, in close vicinity of the PS bead. This reflection then samples the bead in a FWS geometry giving rise to a complex mixture of superimposed FWS and BWS with different driving amplitudes that will not be discussed in further detail. In contrast, FWS and IFS yield straight-forward to rationalise signals whose temporal phase profiles and images match closely, as expected given that IFS is essentially FWS with an internally generated illumination wave.

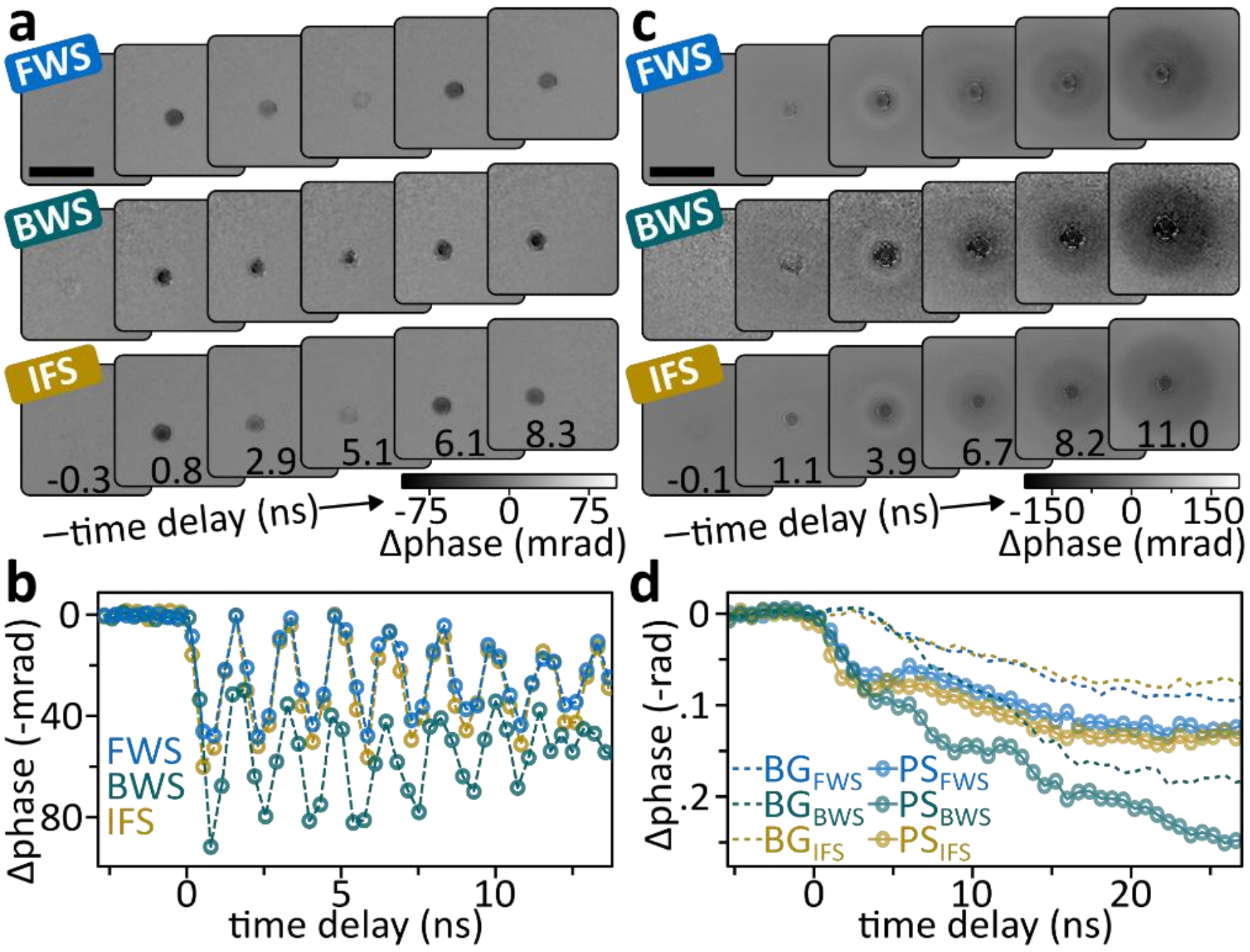

**Figure 4. Photothermal phase dynamics of a 5 µm polystyrene bead in air and water across detection geometries**. a) Representative, time delay dependent, differential phase images for a PS bead in air acquired in FWS, BWS and IFS, showing consistent oscillatory responses. Scalebar: 20 µm. The numbers indicate the time delay in nanoseconds. b) Mean differential phase signals for all three modalities as averaged over the area of the PS bead. c) Representative, time delay dependent, differential phase images for a PS bead in $H_2O$ acquired in FWS, BWS and IFS, showing consistent oscillatory responses. Scalebar: 20 µm. The numbers indicate the time delay in nanoseconds. d) Mean differential phase signals for all three modalities as averaged over the area of the PS bead (solid lines, circles) and a background (BG) region in close vicinity to the bead reporting on the $H_2O$ signal (dashed lines).

Figure 4c summarises the changes observed when moving from a bead contained in air to a water environment. Notably, we now observed differential phase signals which we associated with contributions from both the bead as well as the surrounding, and strongly MIR absorbing, water. Qualitatively, we observed a rapidly rising differential signal at the location of the PS sphere as well as a slowly rising, featureless, background signal, consistent with thermoelastic expansion propagating through the medium at velocities comparable to the speed of sound in water. To gain further insight into bead and background contributions, we compared the temporal dynamics measured at the centre of the bead with a location in close proximity thus providing reference-like dynamics without attempting a potentially artefact-prone decomposition into PS bead and water-background signals (Figure 4d). In line with our previous observations, we noted a consistent behaviour across modalities where FWS and IFS produced near-identical phase responses and dynamics, while BWS yielded the largest overall signal. These observations support the central conclusion that IFS emulates FWS even under strong background absorption.

As mentioned above, BWS observation of a structure near an interface yields a complex mixture of scattering and reflection contributions. An immediate question is what would happen if the structure would not be located near said interface but in an isotropic medium. To approximate such a three-dimensional environment, we performed experiments in a glycerol–water mixture with a refractive index matched to that of the $CaF_2$ substrate thus optically eliminating the substrate and hence the reflection (Figure 5a). Figure 5b compares differential phase changes acquired via IFS and BWS, as accessed by temporal coherence gating by simply adjusting the probe-reference time delay while retaining the overall experimental configuration (Figure 2c). In IFS, phase changes due to both the PS bead and the surrounding solution are apparent, with the background signal being dramatically increased compared to the PS bead in water-only experiments (Figure 4c, 5b). In stark contrast, BWS shows no background signal but a very strong, clover-leaf shaped particle signal which can be qualitatively rationalised by considering Mie scattering (Figure 1b). To rationalise the origin of this signal, we compared the temporal dynamics obtained for IFS and BWS (Figure 5c). Both IFS and BWS show comparable dynamics with similar signal magnitudes which are, however, strongly dependent on the precise signal location given the spatially highly non-uniform BWS signal at the PS bead locations (Figure 5b). Irrespective, on long time-scales the dynamics at the particle locations qualitatively matches those of IFS-background signals obtained at locations in close proximity. On short time-scales deviations between particle and background signals are apparent. The beads exhibit two temporal components which we attribute to a rapidly rising bead signal plus the slowly increasing solvent background. BWS shows essentially no background signal (Figure 5b), which is unsurprising given that the refractive index matched nature of the experiment makes it highly non trivial to obtain any information in these particle-free areas.

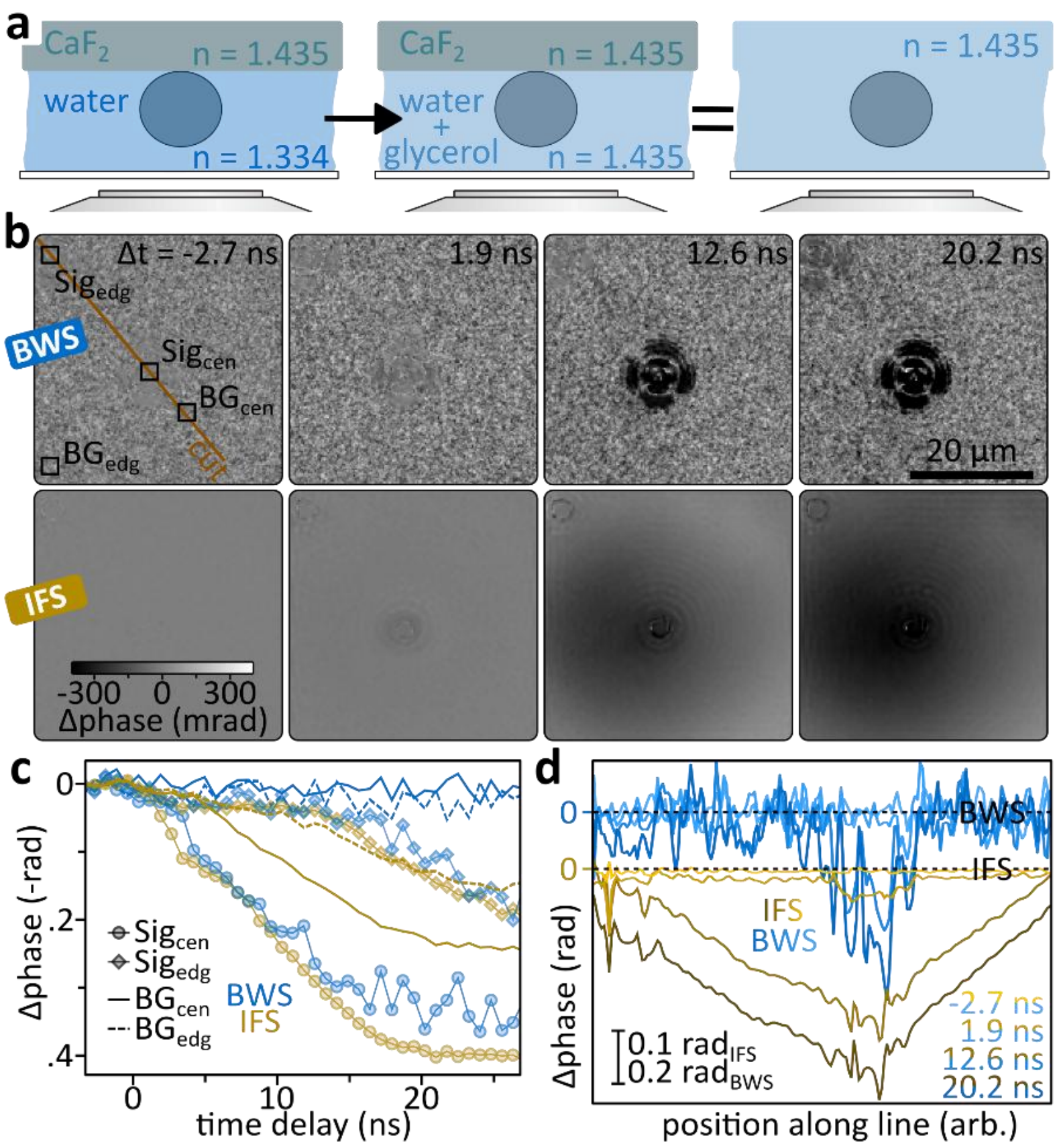

**Figure 5. Impact of illumination geometry in quasi-3D environments.** a) Transitioning from water to a glycerol-water mixture refractive index matches the $CaF_2$ substrate thus suppressing substrate reflections. b) Differential phase imaging at different pump-probe delays, *Δt*, of two 5 µm PS beads at an index-matched $CaF_2$/glycerol-water interface, comparing BWS (top) and IFS (bottom). The squares and the line indicate regions of signal averaging for the temporal dynamics (squares) and the spatial evolution (line). c) Comparison of the temporal evolution of the differential phase signals for BWS and IFS at the previously indicated locations. d) Comparison of the spatial profiles obtained at selected time-delays, as indicated in b). BWS and IFS signals are offset for clarity, as indicated. IFS and BWS signals are scaled by a factor of two with respect to each other, as indicated by the scale bar.

To gain further insight, we compared the temporal evolution along a line profile (Figure 5d) passing through the centre of both PS beads as indicated in Figure 5b. For IFS we observe a rapidly rising bead signal as well as the slow emergence of a broad solvent background. In contrast, BWS shows the emergence of the strongly spatially modulated bead signal with a rise comparable to the solvent background and roughly twice its differential phase change. In empty regions, no signal emerges in line with the explanation given above. The origin of these strikingly different spatial observations (Figure 5b,d) is the combination of bead and solvent signals. In IFS, the transmitted probe directly measures the pump-induced differential phase thus picking up the spatially localised bead signals alongside the broad background in a quantitative manner. In contrast, BWS reports on a combination of solvent background and bead signals through back-scattered probe photons. As the probe field samples the volume twice, traveling to the bead and then back, the resulting BWS signals are roughly twice the IFS signals' magnitude.

Given the combined observations (Figure 3-5) we believe that FWS or IFS are the most suitable observation geometries with IFS being the experimentally easier-to-implement variant. Our rational is simple, BWS might, under certain geometries, yield larger signal magnitudes but this minor boost in signal comes at a huge cost. Firstly, the strong dependence of scattering signals on the sample shape (Figure 1b) is a huge problem that dramatically complicates quantitative measurements. Secondly, in solvent-environments it is generally necessary to interrogate the sample of interest near a MIR-transparent substrate. Even though the interface provides a back-reflection that somewhat mitigates the Mie-scattering problem outlined above, the resulting signals are generally a complex combination of various scattering and reflection contributions (Figures 3,4). Thirdly, eliminating the interface removes complexity by eliminating all but the sample-intrinsic backscattering signal but this aspect results in highly complex spatial signals even for intrinsically simple objects such as uniform spheres (Figure 5). To make matters worse, the back-scattered signals report on background signals as well as those from the object of interest. While generally the case (Figures 3-5), a quantitative spatial measurement of signal and background signals opens straight-forward avenues towards separating the two by, for example, high-pass filtering or similar. BWS provides no information on the underlying background signal as it only measures at positions where particles are present thus rendering background-estimation a highly complex problem.

**SUMMARY & CONCLUSION**

To summarise, we introduced and systematically evaluated a bidirectional femtosecond MIR pump-probe off-axis holographic microscope capable of operating in FWS and BWS and then used this microscope to implement and validate a favourable illumination scheme that we term IFS. IFS is essentially a FWS observation but measured while illuminating the sample in a much-easier to implement geometry that is normally used for BWS. IFS is enabled by the back-reflection of a substrate interface that is further away from the sample than the temporal coherence length of the illumination source and is thus distinct from recently reported MIP imaging schemes relying on specifically tailored reflective substrates[11] or tomographic reflection-acquisition[12]. Time-gating the IFS hologram acquisition through temporal coherence gating then allows isolating the IFS information. By rapidly switching between IFS, BWS and FWS geometries while preserving pump–probe alignment, we directly and quantitatively compared the photothermal phase sensitivity and image quality across all three geometries. To ensure that IFS is broadly applicable, we specifically chose very large phase objects in the form of 5 µm PS beads, roughly matching MIR-penetration in water, that can be regarded as a "worst case scenario" in terms of optical distortion of the incoming illumination light in the IFS configuration.

Our observations are consistent with the expectation that conventional BWS measurements experience reduced structural fidelity and interpretability in the Mie scattering regime, potentially limiting their usefulness for quantitative phase imaging despite their higher intrinsic signal amplitude. In contrast, FWS preserves quantitative phase information but requires two-sided optical access, complicating high-NA implementations for MIR photothermal imaging. The IFS addresses this limitation by combining FWS-like phase sensitivity and closely matched signals with single-sided access, achieved through coherent temporal gating that selectively isolates a forward-scattered component within the nominal BWS geometry.

Across multiple experimental conditions, PS beads in air, water, and refractive-index-matched environments, IFS consistently reproduced the amplitude and temporal dynamics of FWS phase signals while retaining the mechanical simplicity and single-sided accessibility of BWS configurations. Furthermore, systematic measurements of the through-focus profiles comparing IFS, FWS and BWS quantitatively underlined that the former two are essentially identical and significantly differ from BWS (Supplementary Information 3). Measurements in index-matched media further confirmed the existence of two distinct BWS contributions and validated the temporal gating strategy underlying IFS. Together, these findings highlight the promise of IFS as an alternative imaging modality for quantitative MIR photothermal microscopy and holographic microscopies in general. Importantly, BWS remains the method of choice for nanoscopy applications in the Rayleigh regime where near isotropic signals, combined with the crucially important illumination light attenuation through reflection, offer clear advantages. In terms of applicability, IFS can readily be implemented using off-the-shelf diode lasers which exhibit temporal coherence lengths around 100-200 µm thus enabling the coherence gating strategies proposed here[24].

The demonstrated equivalence between FWS and IFS opens new opportunities for high-resolution, one-sided MIR photothermal imaging in experimentally constrained settings. Biological samples are a prime candidate where the need for high resolution combined with aqueous environments as well as temperature control dramatically complicate photothermal as well as conventional IR-imaging. We will extend the IFS approach towards biologically relevant specimens, including live cells and tissues that are currently imaged via FWS-schemes, to assess its robustness under complex refractive-index landscapes. Integrating high-NA objectives and faster delay or modulation schemes could further improve spatial and temporal resolution. More broadly, the IFS concept may be adaptable to other pump–probe or quantitative phase imaging modalities where access geometry plays a critical role.

## METHODS

### Light Generation

Spectrally tuneable MIR pump pulses were produced using a home-built multi-stage noncollinear optical parametric amplifier (NOPA) pumped by the second harmonic (515 nm) of an amplified Yb:KGW laser (Pharos PH2-20W-SP, Light Conversion; 1 kHz, 2 mJ, 800 fs). A white-light continuum was generated by focusing <1 µJ of the 1030 nm fundamental into a 10 mm YAG crystal and subsequently amplified in a double-pass noncollinear optical parametric amplifier (noncollinear-OPA) tuned for broadband phase matching, yielding 2–3 µJ of tuneable light (650–980 nm, ~150 $cm^{-1}$ bandwidth). An OPA booster stage comprising three BBO crystals[28], pumped with 500 µJ of 515 nm light, further amplified this seed to produce MIR pulses tuneable from 1150–1850 nm. MIR pulses of 1593 $cm^{-1}$ (~ 8 µJ) where generated via difference frequency generation in a type II AGS crystal by combining the chirped, ~ 3 ps, OPA output ( ~ 40 µJ) with residual, equally chirped, 1030 nm light of the fundamental (~ 1 mJ). MIR light was spatially separated from the two driving beams by operating the difference frequency generation in a slightly non-collinear geometry. A small fraction of the 1030 nm fundamental was frequency-doubled to generate the 515 nm probe pulse. The MIR spectra were characterized using a home-built Michelson interferometer together with a MIR-photodiode.

### Microscope Setup

The 515 nm probe beam was split into sample and reference arms using a 50/50 beamsplitter. A mirror mounted on a linear translation slider enabled rapid switching between FWS and BWS/IFS detection geometries without disturbing MIR pump alignment. For FWS, the MIR pump and the 515 nm probe were collinearly combined by reflecting the 515 nm probe off a Ge window which transmitted the MIR pump. The MIR pump (<4 µJ, 1593 $cm^{-1}$) was modulated at 500 Hz using a mechanical chopper and focused onto the sample with a 40 mm Ge lens. Probe images were acquired at 1 kHz, producing alternating pump-on/pump-off holograms for differential phase retrieval.

Scattered probe light was collected using a 20×/0.45 NA objective. A 200 mm tube lens formed an image on a CMOS camera (Basler acA1920-155um) at an effective magnification of 22×, corresponding to an approximately 41.1 µm × 41.1 µm field of view. Probe–reference delay matching was performed using a manual translation stage, while pump–probe delay scans were controlled using two motorized stages: a fine-stepping stage (M230.25, Physik Instrumente) and a long-travel stage (FPB45, 2000 mm, FUYU Automation) operated in a double-pass configuration and driven by an in-house Nexys A7-100T–based controller. Active beam stabilization ensured drift-free acquisition for mechanical delays exceeding 7 m.

**Data Processing**

Images were acquired at 1000 frames per second and an exposure time of 40 microseconds per frame, with equal numbers of pump-on and pump-off frames. Each hologram was Fourier-transformed, and a single off-axis interference term was isolated in k-space, shifted to the origin, and inverse-transformed to recover the complex field (see Figure 2d). Hot and cold complex images were separated, and the differential phase was extracted from the ratio of the two complex fields and subsequently all 500 differential phases were averaged. The same process was applied consistently across all the geometries.

**Sample Preparation**

Polystyrene beads were diluted in absolute ethanol, deposited onto $CaF_2$ windows, and allowed to dry, thus the sample is static. #1.5H cover glasses were cleaned by 20 min sonication in acetone followed by 20 min in Milli-Q water. The cover glass was fixed to a 1 mm $CaF_2$ window using 200 µm double-sided tape, forming a shallow chamber for introducing aqueous or glycerol/water solutions. The sample was tilted by <1° to avoid camera saturation due to back reflections.

**Refractive-Index-Matching Solution**

Refractive-index-matched media were prepared by using the Lorentz–Lorenz mixing rule to calculate the fractional volumes of water and glycerol required to obtain a target effective refractive index matching that of the $CaF_2$ substrate. Based on this calculation, a mixture of 0.31 mL water and 0.69 mL glycerol was prepared to achieve an effective refractive index of $n_{eff}$ = 1.433.

## Supplementary Information:

## Bidirectional phase sensitivity in holographic phototransient microscopy

Emmanuel Kotu Robertson[1], Dennis van de Lockand[1], Matz Liebel[1,a]

[1] Department of Physics and Astronomy, Vrije Universiteit Amsterdam, De Boelelaan 1100, Amsterdam 1081 HZ, The Netherlands

[a] email: m.liebel@vu.nl

### Supplementary information 1 – Plane wave recovery after back-reflection and transmission through substrate

To showcase that the imprinted phase profile acquired during the first pass through the sample rapidly disappears, we simulated a 5 µm PS (n=1.59) bead in air as a pure, flat, phase object with a spherical phase profile according to its z-projected thickness (peak phase retardation of 36 rad at an illumination wavelength of 515 nm). We then used the angular spectrum method to propagate the resulting field through $CaF_2$ with a thickness of 340 µm and 2000 µm, representing a "typical coverglass" and the substrate thickness used in our work (two times 1000 µm), respectively. Figure S1 shows that an essentially flat wavefront is recovered after 2000 µm with only a minor residual phase imprint being present after 340 µm. Importantly, even the 340 µm phase imprint is smaller than most "phase imperfections" in typical "plane wave" illumination sources as dust, speckle etc. generally impact the illumination wave.

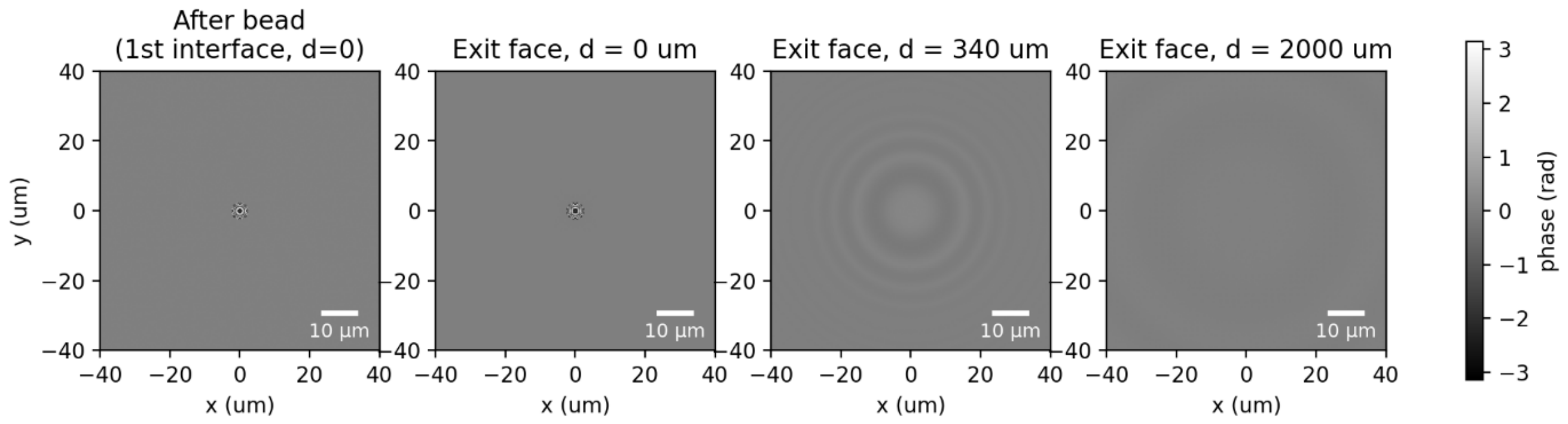

**Figure S1.** Transverse phase (wrapped, radians) immediately after the bead (left) and at the substrate exit face for d = 0, 340, and 2000 µm. The concentric-ring pattern imprinted by the bead spreads and dilutes with increasing thickness; by d = 2000 µm the field within the plotted window (±40 µm) is visually uniform.

### Supplementary information 2 - Coherence gating characterization

To validate the path selectivity of the temporal coherence gate, we recorded the reconstructed interference amplitude versus reference-arm delay across the sample chamber depth (Figure S2A). Five peaks are resolved and assigned to the near-interface BWS field (B), the IFS channel generated by back-reflection at the distal surface of the $CaF_2$ substrate (A), the front and back surfaces of the coverglass (C, D), and an additional weak reflection (E). Peak-to-peak spacings closely track the expected optical path lengths for each interface: the A–B separation (1.549 mm) is consistent with a double pass through the 1.06-mm-thick $CaF_2$ substrate (n = 1.433 at 515 nm; expected 1.519 mm, ~2% agreement), confirming the assignment of peak A to the distal-surface-reflected IFS field. The B–C spacing (0.197 mm) matches the ≈0.2 mm air gap between the substrate and coverglass, and the C–D spacing (0.269 mm) matches a double pass through the 0.17-mm-thick coverglass ($n \approx 1.58$ at 515 nm). All measured peak widths (FWHM: A = 0.062 mm, B = 0.060 mm, C = 0.070 mm, D = 0.080 mm, E = 0.115 mm) are comparable to the coherence length expected from the 515 nm source bandwidth ($\Delta\lambda \approx 3$ nm, $L_c \approx \lambda^2/\Delta\lambda \approx 88$ µm), and are much narrower than the ≥0.2 mm separation between any two adjacent peaks. The IFS and BWS channels are therefore well resolved by the coherence gate, with negligible cross-talk between them.

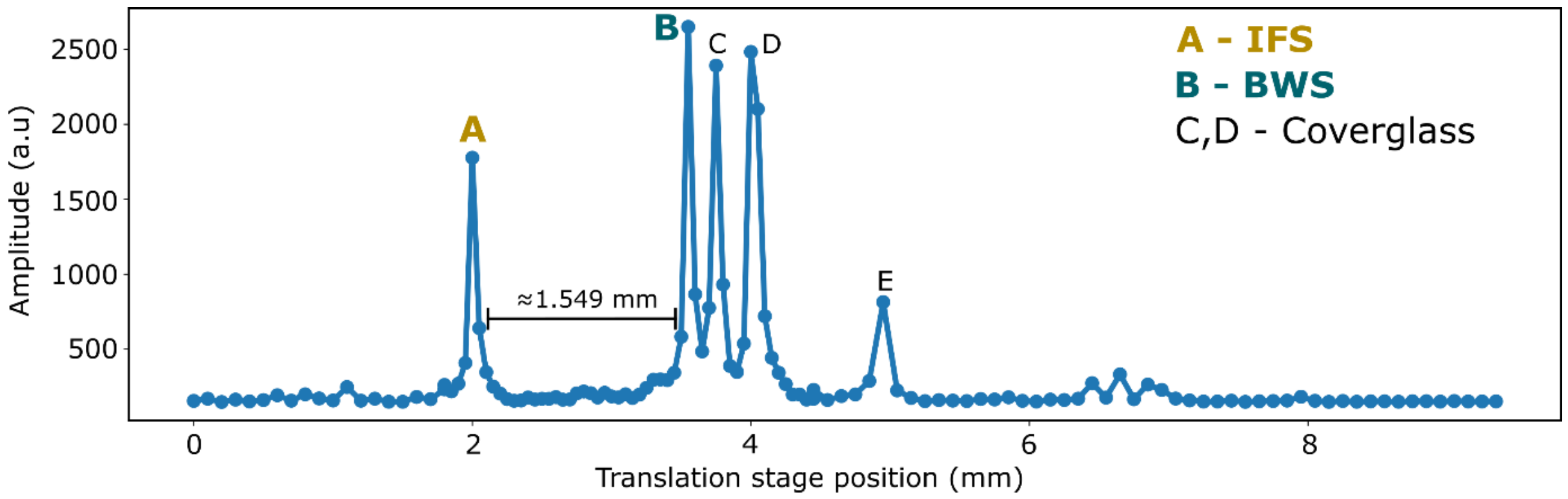

**Figure S2A.** Reference delay scan in BWS geometry.

As a control for parasitic reflections independent of the sample interfaces, we performed the same delay scan in the forward-scattering (FWS) configuration (Figure S2B). A single dominant FWS peak is observed (FWHM = 0.062), along with one weak secondary peak at a delay matching peak E in Figure S2A. As this feature appears in both configurations at the same delay and is well separated from the FWS, IFS, and BWS peaks used for quantitative analysis, we attribute it to a parasitic reflection from a microscope optic (e.g., the objective) rather than to the sample chamber or particle, confirming it does not contaminate the channels of interest.

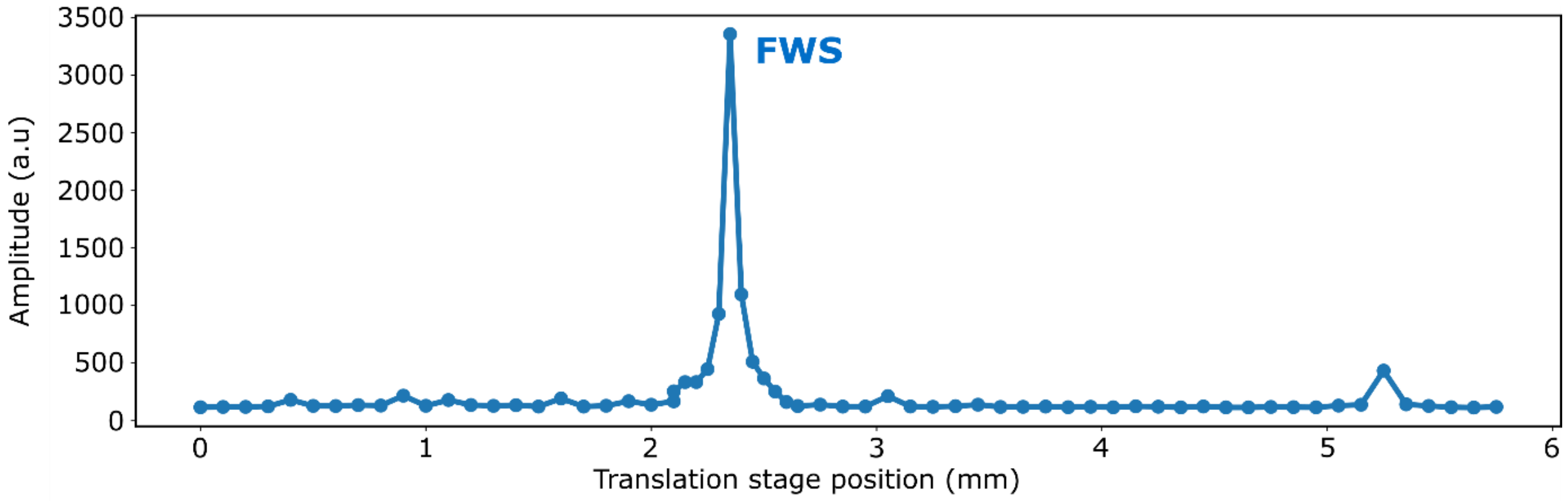

**Figure S2B.** Reference delay scan in FWS geometry.

**Supplementary information 3 - Quantitative comparison of static bead phase across FWS, BWS and IFS**

We imaged many individual 1 µm diameter polystyrene spheres on glass and propagated their fields via the angular spectrum method[1] to obtain through-focus signal evolution, indirectly accessing their Mie-scattering profile. Comparing this evolution (rather than absolute z-focus) also eliminates variations from shifting absolute focal positions between images. As shown in Figure S3 (mean ± std across all imaged spheres), FWS and IFS profiles are essentially identical, while BWS diverges significantly. The minor FWS/IFS differences likely stem from small misalignments in illumination k-vectors — e.g., a slight tilt relative to normal incidence (0°) in one of the two.

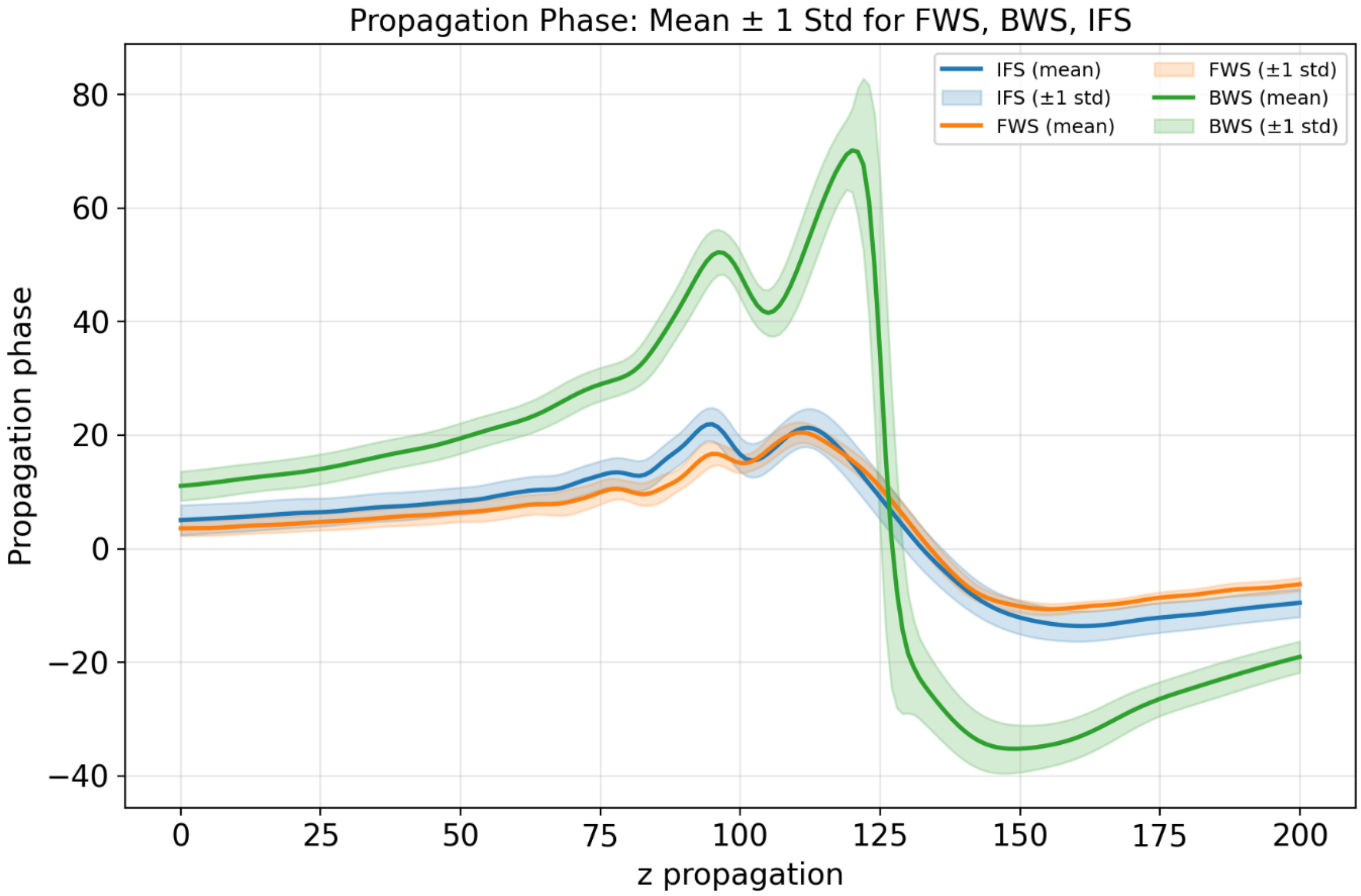

**Figure S3.** Quantitative comparison of IFS, FWS, and BWS signals, and focus-independence check.